\documentclass[conference]{IEEEtran}
\usepackage[utf8]{inputenc}
\usepackage{cite}
\usepackage{floatrow}
\usepackage{amsmath,amssymb,amsfonts,stfloats}
\usepackage{tabularx}
\usepackage{algorithmic}
\usepackage{graphicx}
\usepackage{textcomp}
\usepackage{floatrow}
\usepackage{xcolor}
\usepackage{booktabs}

\usepackage{multicol}
\usepackage{acronym}
\usepackage{url}
\usepackage{mathtools}
\usepackage[caption=false,font=footnotesize]{subfig}

\def\BibTeX{{\rm B\kern-.05em{\sc i\kern-.025em b}\kern-.08em
    T\kern-.1667em\lower.7ex\hbox{E}\kern-.125emX}}

\begin{document}
 
\title{Dual Domain Waveform Design for \\Joint Communication and Sensing Systems}

\author{\IEEEauthorblockN{Luca Rinaldi, Dario Tagliaferri, Francesco Linsalata, Marouan Mizmizi, \\ Maurizio Magarini and Umberto Spagnolini}
\IEEEauthorblockA{Dipartimento di Elettronica, Informazione e Bioingegneria, Politecnico di Milano, Via Ponzio 34/5, 20133, Milano
Italy}
E-mails: \{luca.rinaldi,dario.tagliaferri,francesco.linsalata,marouan.mizmizi,maurizio.magarini,umberto.spagnolini\}@polimi.it}

\maketitle

\begin{abstract}
The evolution of wireless communication systems towards millimeter-wave ($30-100$ GHz) and sub-THz ($>100$ GHz) frequency bands highlighted the need for accurate and fast beam management and a proactive link-blockage prediction in high-mobility scenarios. Joint Communication and Sensing (JC\&S) systems aim at equipping communication terminals with sensing capabilities using the same time/frequency/space communication resources to solve, or alleviate, the aforementioned issues. For an efficient implementation, a suitable waveform design that combines communication and sensing capabilities is of utmost importance. This paper proposes a novel dual-domain waveform design approach that superimposes onto the Frequency-Time (FT) domain both the legacy orthogonal frequency division multiplexing modulation scheme and a sensing signal, purposely designed in the Delay-Doppler (DD) domain. The power of the two signals is properly allocated in FT and DD domains, respectively, to reduce their mutual interference and optimize both communication and sensing tasks. 
Numerical results show the effectiveness of the proposed JC\&S waveform design approach, yielding target communication and sensing performance with a full time-frequency resource sharing.
\end{abstract}


\maketitle

\section{Introduction}\label{Introduction}

JC\&S is emerging as one of the key technologies for the upcoming 6th Generation (6G) of communication systems for advanced applications in the Millimeter-Wave (mmWave) ($30-100$ GHz) and sub-THz ($> 100$ GHz) bands~\cite{rahman2020enabling,Heath2021overview,wild2021JCS6G,Prelcic2020leveraging}.
This new paradigm has been attracting a major interest in the scientific community, since it combines two traditionally independent functionalities, i.e.,  communication and sensing, into a single one, thus sharing the same frequency/time/space resources and hardware. 

The design of a single suitable waveform is the most critical challenge in JC\&S systems, as the communication and sensing functionalities call for different waveform requirements~\cite{Heath2021overview}. The most general approach is to design the waveform by optimizing the performance of one functionality leaving the other one constrained by a suitable metric~\cite{liu2018toward}. 
Orthogonal Frequency Division Multiplexing (OFDM) was the first considered waveform for JC\&S, where specific symbol-based processing has made possible radar measurements~\cite{Sturm2011}. Accurate range resolution was demonstrated in a single-user scenario by considering the entire available bandwidth. However, the power of the transmitted signal is tailored to the communication with the users, and, in a real scenario, the back-scattered OFDM signal used for sensing would experience an orders-of-magnitude increase in the path-loss (two-way propagation), resulting in a very low Signal-to-Noise Ratio (SNR). More recently, in~\cite{Barneto2021_Optimized_Wave}, the authors propose to fill empty subcarriers in the conventional time-frequency OFDM resource grid and free from communication, with sensing pilots (i.e., radar subcarriers). The proposed method dynamically optimizes both the power and the phase of the radar subcarriers, aiming at minimizing the Cram\'{e}r-Rao bound on range and Doppler estimation while limiting the peak-to-average power ratio. However, the effective sensing performance is ruled by the communication load. Moreover, the optimization has to be repeated whenever the communication load changes, increasing the complexity of the implementation. A different contribution is presented in~\cite{Heath2020_Virtual}, where the authors propose a single-carrier mmWave waveform structure in which the preamble of the communication frame is dynamically exploited as a radar pulse.

Orthogonal Time Frequency Space (OTFS) modulation has been recently receiving increased interest as a good candidate waveform for JC\&S thanks to its robustness in high Doppler environments~\cite{hadani2018orthogonal}. 
OTFS has shown to have better performance compared in doubly-selective channels compared to OFDM and allows to achieve sensing performance of the same order of latest radar waveforms~\cite{gaudio2019performance}. However, it requires a substantial modification of the current 5th Generation (5G) New Radio (NR) standard since it needs processing bursts of consecutive OFDM symbols, which increases the latency and the computational burden, being it in contrast with the demand for low latency services in 6G. 

Alternative solutions consider the design of an optimal beam pattern that is capable of delivering information data with one beam and sensing the environment with another one. The authors in~\cite{liu2018beampattern} optimize the communication beamformer such that the achieved beam pattern matches the radar’s one while satisfying the communication performance requirements. The optimal waveform is designed with an explicit Pareto optimization, that gives fixed weights to both multi-user interference and optimal radar beam pattern.

In this paper, we aim at filling the aforementioned gaps in the literature, proposing a JC\&S system at the infrastructure, e.g., a Base Station (BS), enabling a dual-functionality: communication transceiver for User Equipments (UEs) and monostatic radar. In this context, we propose a novel dual domain JC\&S waveform combining a sensing signal, designed in the Delay-Doppler (DD) domain, with the legacy OFDM signal, designed in the Frequency-Time (FT) domain. The sensing signal is first converted to the FT domain and then fully superimposed to OFDM. The advantage of the proposed approach is that, with proper sensing parameters' design, the mutual interference between communication and sensing signals can be tuned to meet target performance without an explicit trade-off, with a complete sharing of the time-frequency resources between the two functionalities. Numerical simulations are presented to show the benefits of the proposed JC\&S system.

The remainder of the paper is organized as follows: Section \ref{Time-variant Scenario} outlines the considered scenario and the channel model, Sec. \ref{ProposedWaveform} describes the proposed JC\&S waveform and related processing, Sec. \ref{Simulations} outlines the simulation results while Sec. \ref{Conclusion} draws the conclusion.
\begin{figure*}[!t] 
\centering
\subfloat[][]{\includegraphics[width=0.45\columnwidth]{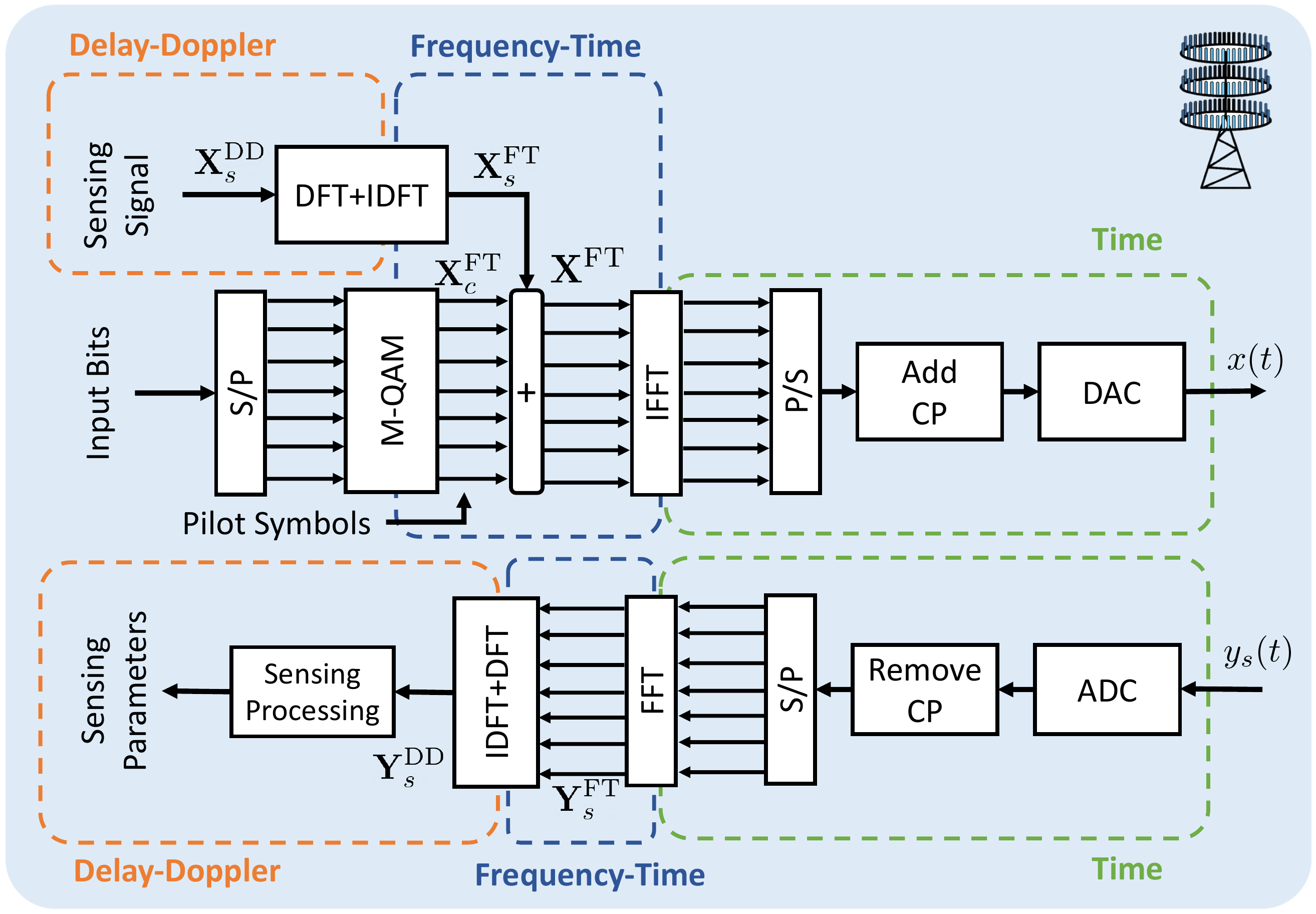}\label{fig:JCAS_TX}}
\subfloat[][]{\includegraphics[width=0.45\columnwidth]{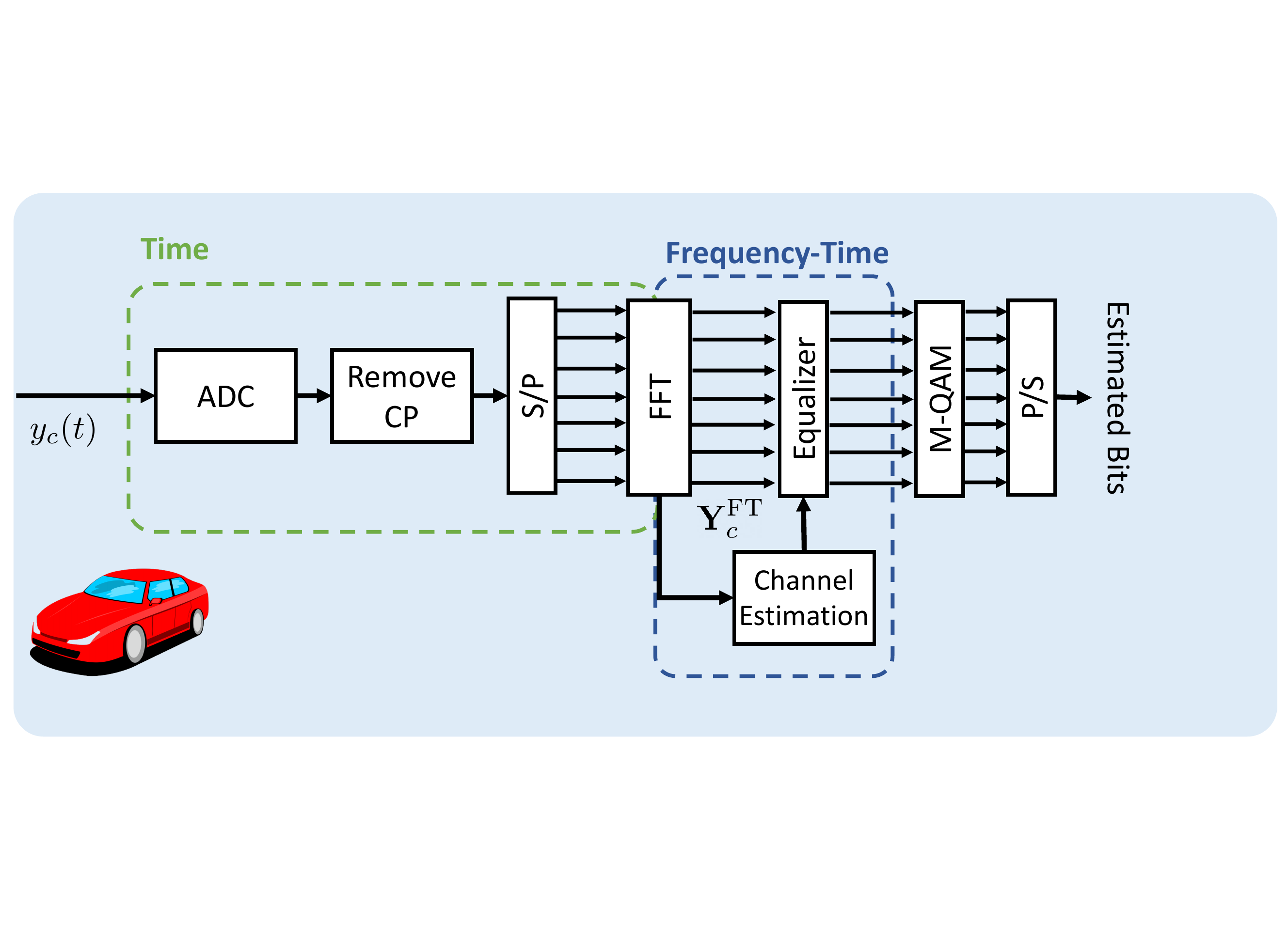}\label{fig:JCAS_RXs}}
\caption{Block scheme of the proposed JC\&S system: (a) JC\&S Tx (Top-left) and sensing Rx (Bottom-left) (BS); (b) Communication receiver (UE)}
\label{fig:System Model}
\end{figure*}
\section{Time-Variant Scenario}\label{Time-variant Scenario}
%
The scenario we are considering in this work is a cellular communication system where a BS serves $U$ UEs, randomly located in space, through the legacy OFDM waveform, while sensing the environment as a monostatic radar. For the purpose of this work we consider a Single-Input Single-Output (SISO) system. The proposed single beam processing can be extended to a multi-beam processing with proper generalizations and taking into account also the spatial domain. 

In high-mobility scenarios, such as the vehicular one, the channel is highly variable and characterized by severe path loss and large Doppler frequency shifts due to the UEs' motion.
Specifically, the vehicular channel is doubly (time-frequency) selective, and the time-varying impulse response between the BS and the $u$-th UE is modeled as the sum of $Q_u$ paths as~\cite{samimi2016mmwave}
\begin{equation}\label{eq:channel_model_TD_comm}
    h_{u}(t,\tau) = \sum_{q=1}^{Q_u} \alpha_{q,u} \, e^{j 2 \pi \nu_{q,u} t} \, g(\tau-\tau_{q,u})
\end{equation}
where: \textit{(i)} the amplitude of the $p$-th path is a random circularly complex Gaussian variable, $\alpha_{q,u}\sim\mathcal{CN}(0,\Omega_{q,u})$, with a variance inversely proportional to the square of the distance $R_{q,u}$, i.e., $\Omega_{q,u}\propto R_{q,u}^{-2}$; \textit{(ii)} $\nu_{q,u} = (f_0/c) V_{u}\cos\psi_{q,u}$ is the Doppler shift associated with the velocity of the $u$-th UE, $V_{u}$, where $\psi_{q,u}$ is the angle between the UE's planar direction of motion and the $q$-th path angle of arrival; \textit{(iii)} $g(\tau - \tau_{q,u})$ is the pulse shaping filter, delayed by the $q$-th path delay $\tau_{q,u}$. In a Line-of-Sight (LOS) propagation environment, $R_{1,u} = R_u$ is the \textit{range} of the UE from the BS. Note that \eqref{eq:channel_model_TD_comm} describes the channel in the Time-Delay (TD) domain that is referred to as \textit{slow time-fast time} in the radar jargon.

To model the two-way (BS-UE/target-BS) sensing channel, we make the assumption that all the radar targets are UEs to be served. 
Moreover, as we are considering a monostatic sensing at the BS, we neglect the multi-path in the received echoes, that are considerably weaker than the LoS ones~\cite{Folster2005}.
In this setting, the sensing channel is modeled as 
\begin{equation}\label{eq:channel_model_sensing}
    h_s(t,\tau) = \sum_{u=1}^{U} \alpha_{u} \, e^{j 2 \pi \nu_{u} t} \, g(\tau-\tau_{u})
\end{equation}
where $\alpha_{u}\sim\mathcal{CN}(0,\Omega_{u})$, with $\Omega_{u}\propto R_{u}^{-4}$ is now the scattering amplitude of each UE and 
\begin{equation}\label{eq:delay_doppler sen}
        \tau_{u} = \frac{2R_{u}}{c},\,\,\,\,
        \nu_{u} = \frac{2f_0}{c} V_{u} \cos\psi_{u}
\end{equation}
are the $u$-th UE two-way delay and Doppler shift, respectively.

\textit{Remark}: Considering all the radar target as UEs does not limit the generality of the proposed approach. In general, radar sensing deals with a much higher number of targets than the number of served UEs $U$, and a target association procedure in needed, although not covered here, to discriminate between passive targets and active targets.
\section{Proposed JC\&S Waveform Design}\label{ProposedWaveform}
The transmitted JC\&S signal is designed as the combination of two waveforms, one for communication and the other for sensing, as depicted Fig.~\ref{fig:System Model}. The resulting waveform can be represented on a discrete FT grid, defined as:
\begin{equation}\label{eq:TFgrid}
    \begin{split}
        \Lambda^{\mathrm{FT}} = \{m\Delta f, nT \},\,\,&m =0 ,...,M-1,\\
               & n = 0, ...,N-1, 
    \end{split}
\end{equation}
where $\Delta f$ and $T$ are the subcarrier spacing and the OFDM symbol duration, respectively, being $M$ and $N$ the number of system subcarriers and the number of consecutive symbols in a downlink burst. 
The BS allocates a portion of the FT resources to the communication to each UE as
\begin{equation}\label{eq:TFgrid_user}
    \begin{split}
        \Lambda_{c,u}^{\mathrm{FT}}= \{m\Delta f, nT \}\subseteq \Lambda^{\mathrm{FT}} ,\,\,&m \in \mathcal{M}_{c,u},\,\,
                n\in \mathcal{N}_{c,u}, 
    \end{split}
\end{equation}
where $\mathcal{M}_{c,u}$ and $\mathcal{N}_{c,u}$ are the $u$-th user contiguous subsets of allocated subcarriers and OFDM symbols, of cardinality $M_{c,u} \leq M$ and $N_{c,u} \leq N$. The communication bandwidth dedicated to the $u$-th UE is $B_{c,u} = M_{c,u}\Delta f$. For the design of sensing waveform, we assume that the $u$-th UE communication FT grid $\Lambda_{c,u}^{\mathrm{FT}}$ is randomly placed in the whole $\Lambda^{\mathrm{FT}}$ grid. This mimics a dynamic resource allocation for communication by the BS, and it does not overlap with other communication resources, i.e., $\Lambda_{c,u}^{\mathrm{FT}}\cup\Lambda_{c,\kappa}^{\mathrm{FT}} = \emptyset$, $\forall u\neq\kappa$. 

The $u$-th UE communication signal is therefore written as:
\begin{equation}\label{eq:Comm signal UE}
    \begin{split}
        X^{\mathrm{FT}}_{c,u}[m,n] = \begin{dcases}
          \sqrt{P_{c,u}^{\mathrm{FT}}}\,a[m,n] & \text{if} \,m \in \mathcal{M}_{c,u}, n\in \mathcal{N}_{c,u}\\
          \quad 0 & \text{otherwise}
        \end{dcases},
   \end{split}
\end{equation}
where $a[m,n]$ is a random i.i.d. information symbol drawn from a QAM constellation with $\mathbb{E}[\lvert a[m,n]\rvert^2]=1$, and $P_{c,u}^{\mathrm{FT}}$ is the allocated Tx power over each single FT bin. The overall FT communication transmitted signal can be expressed as
\begin{equation}\label{eq:Comm signal total}
    \mathbf{X}^{\mathrm{FT}}_{c} = \sum_{u=1}^{U} \mathbf{X}^{\mathrm{FT}}_{c,u},
\end{equation}
where $\mathbf{X}^{\mathrm{FT}}_{c,u}\in\mathbb{C}^{M\times N}$ denotes the $u$-th UE communication signal FT matrix.

The sensing signal is designed in the FT-reciprocal DD domain
\begin{equation}\label{eq:DDgrid}
    \begin{split}
        \Lambda_s^{\mathrm{DD}} = \{\ell\Delta \tau, k\Delta \nu \},\,\,&\ell =0 ,\dots,M-1,\\
               & k =  0, \dots,N-1, 
    \end{split}
\end{equation}
where the implicit assumption is that the BS allocates all the available resources for sensing. In \eqref{eq:DDgrid}, $\Delta \tau=1/(M \Delta f)$ and $\Delta \nu = 1/(N T)$ are the delay and Doppler resolution of the system, respectively, ruled by the sensing signal bandwidth $B_s = M \Delta f$ and by the duration of the the downlink burst. 

The sensing waveform is a sparse signal in the DD domain composed of $I\ll N M $ random uncorrelated and equi-powered pulses located in $\{[\ell_i,k_i]\}_{i=0}^{I-1}$: 
\begin{equation}\label{eq:Sensing DD matrix multi imp}
     X^{\mathrm{DD}}_{s}[\ell,k] = \begin{dcases}
          \sqrt{P_s^{\mathrm{DD}}} \, s[\ell, k] & \text{if}\,\, \ell=\ell_i, k=k_i\\
          \quad 0 & \text{otherwise}
        \end{dcases}.
\end{equation}
The amplitudes are modelled as $s[\ell, k] \sim\mathcal{CN}(0,1)$ and $P_s^{\mathrm{DD}}$ is the single pulse power.
The BS maps the sensing signal \eqref{eq:Sensing DD matrix multi imp} from the DD domain to the FT domain by~\cite{Bello63}:
%
%
\begin{equation}\label{eq:discreteISFFT}
    \mathbf{X}^{\mathrm{FT}}_s = \mathbf{F}_{M}\mathbf{X}^{\mathrm{DD}}_s\mathbf{F}^{\mathrm{H}}_{N},
\end{equation}
where $\mathbf{F}_{K}$ is the generic $K$-point Discrete Fourier Transform (DFT) matrix (with $K$ = $M$ or $N$). 
Notice that, with \eqref{eq:discreteISFFT}, each pulse in the DD domain maps into a complex 2D sinusoid in the FT domain. The key idea of the proposed method is therefore to allow the superposition of communication and sensing signals in FT, designing the system parameters, namely $\{P_{c,u}^{\mathrm{FT}}, M_{c,u}, N_{c,u}\}_{u=1}^U$ and $P_{s}^{\mathrm{DD}}$, to guarantee a target performance level for both functionalities. For the sake of simplicity, let's assume a single-pulse sensing waveform. By exploiting the intrinsic property of DFTs, the power of the corresponding 2D sinusoid over the TF grid, $\Lambda_s^{\mathrm{FT}}$ = $\Lambda^{\mathrm{FT}}$, is abated by a factor $M N$, and the $u$-th UE communication signal power over the sensing DD grid $\Lambda_s^{\mathrm{DD}}$ is scaled properly 
\begin{equation}\label{eq:power_sensing_FT}
    P_s^{\mathrm{FT}} = \frac{P_s^{\mathrm{DD}}}{M N},\hspace{1cm}P_{c,u}^{\mathrm{DD}} = \frac{P_{c,u}^{\mathrm{FT}} M_{c,u} N_{c,u}}{M N}.
\end{equation}
\begin{figure}[!t] 
    \centering
    \vspace{0.3 cm}\includegraphics[width=7cm, height = 6.5cm]{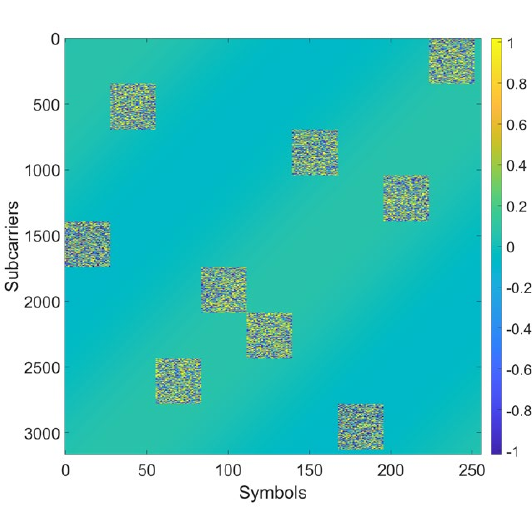}
    \caption{Transmitted JC\&S waveform $\mathbf{X}^{\mathrm{FT}}$, superimposing the OFDM signals of $U=9$ UEs and the sensing signal, a 2D sinusoid covering all the FT grid.}
    \label{TX_FT}
\end{figure}
The overall discrete-time transmitted JC\&S signal, depicted in Fig.~\ref{TX_FT}, is thus the superimposition of the communication and the sensing signals in FT domain:
\begin{equation}\label{eq:JCAS_TX_sig}
    \mathbf{X}^{\mathrm{FT}}=\mathbf{X}^{\mathrm{FT}}_{c} + \mathbf{X}^{\mathrm{FT}}_{s}.
\end{equation}
\begin{figure*}[!t]
    \begin{equation}\label{eq:comm_channel_DD}
        H_{c,u}^{\mathrm{DD}}[\ell,k] = \frac{1}{\sqrt{MN}} \sum_{q=1}^{Q_u} \alpha_{q,u}\,\frac{\sin\left(\pi(\nu_{q,u}NT - k)\right)}{\sin\left(\frac{\pi}{N}(\nu_{q,u}NT - k)\right)} e^{- j \pi \frac{(\nu_{q,u} NT - k)(N-1)}{N}}\,g(\ell T_s - \tau_{q,u})
    \end{equation}
    \hrule
\end{figure*}
\begin{figure}[!t]
\includegraphics[width=7cm]{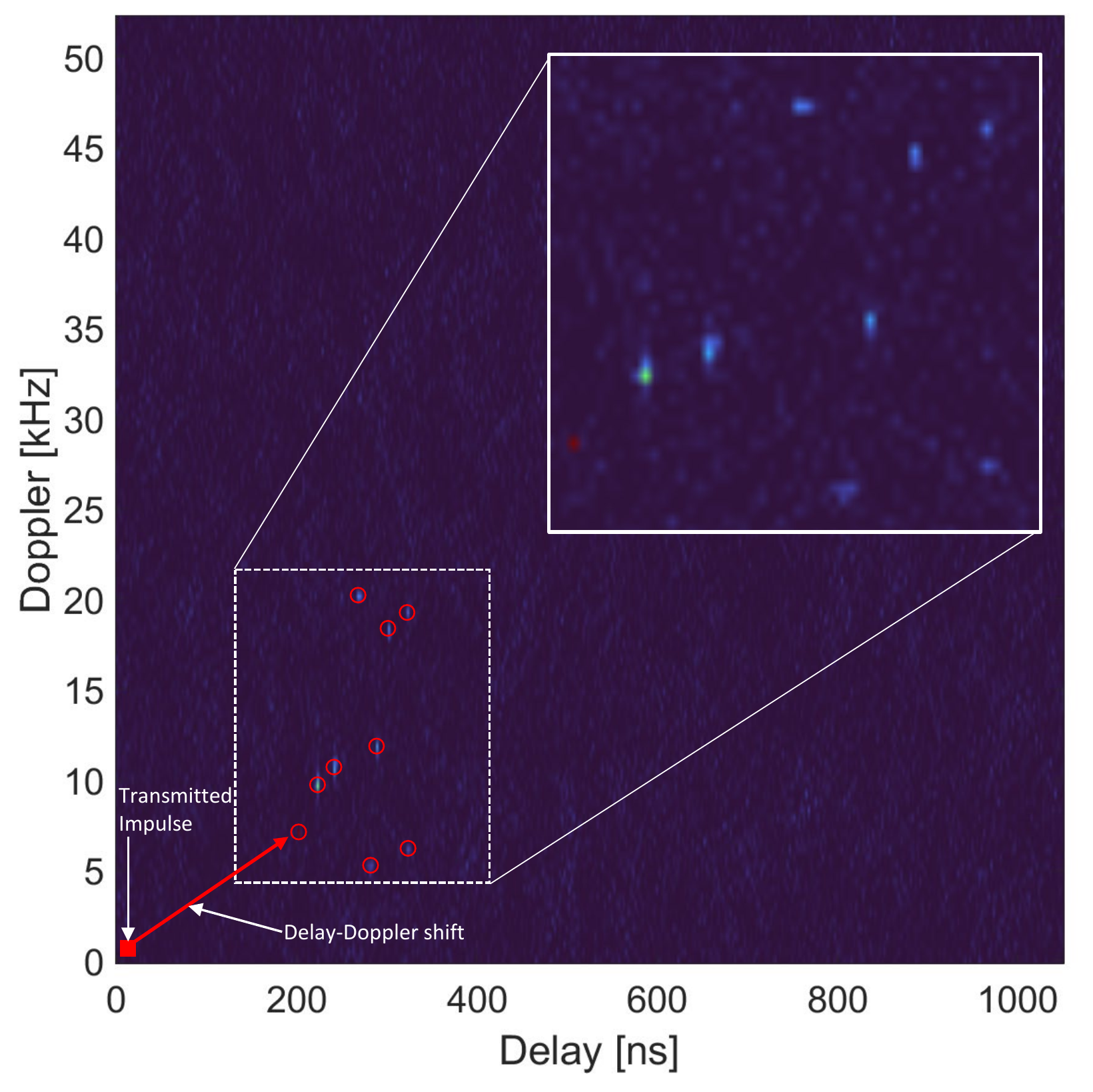} 
\caption{Received signal at the BS in DD domain $\mathbf{Y}_s^\mathrm{DD}$. The transmitted sensing impulse in DD domain is convolved with the DD response of the two-way channel obtaining delay and Doppler representation of each UE.}
\label{RX_DD}
\end{figure}
The transmitted JC\&S signal, $\mathbf{X}^{\mathrm{FT}}$, after the application of the Cyclic Prefix (CP). propagates in the sparse doubly-selective channel, both the communication (BS-UE) and sensing (BS-UE/target-BS). The received communication and sensing signals are (after removing the CP)
\begin{align}\label{eq:TFreceivedsignal_sensing}
    \mathbf{Y}_{c,u}^{\mathrm{FT}} = \mathbf{H}^{\mathrm{FT}}_{c,u}\odot \mathbf{X}^{\mathrm{FT}} + \mathbf{N}_{c,u}^{\mathrm{FT}}\hspace{0.4cm}\text{at UE},\\\label{eq:TFreceivedsignal}
    \mathbf{Y}_s^{\mathrm{FT}} = \mathbf{H}^{\mathrm{FT}}_s\odot \mathbf{X}^{\mathrm{FT}} + \mathbf{N}_s^{\mathrm{FT}}\hspace{0.4cm}\text{at BS},
\end{align}
where $\odot$ denotes the element-wise multiplication while $\mathbf{H}^{\mathrm{FT}}_{c,u}$ and $\mathbf{H}^{\mathrm{FT}}_s$ are the $u$-th UE discrete communication and sensing channels in the FT domain, respectively. $\mathbf{N}_{c,u}^{\mathrm{FT}}$ and $\mathbf{N}_s^{\mathrm{FT}}$ are the noise-plus-interference matrices at the $u$-th UE communication receiver and at the sensing receiver (BS), with elements $N^{\mathrm{FT}}[m,n]\sim\mathcal{CN}(0,P_n)$. 
The communication channel $\mathbf{H}^{\mathrm{FT}}_{c,u}$ is obtained from \eqref{eq:channel_model_TD_comm} by applying a Fourier transform along the delay dimension and sampling in time and frequency,
\begin{equation}\label{eq:discrete FT channel}
    \begin{split}
        H_{c,u}^{\mathrm{FT}}[m,n] =
        \sum_{q=1}^{Q_u} \alpha_{q,u} \,e^{j 2 \pi( \nu_{q,u} nT - m\Delta f \tau_{q,u})} G(m\Delta f),
    \end{split}
\end{equation}
where $G(m\Delta f) = \int g(\tau)e^{- j 2 \pi m\Delta f\tau}\, d\tau$ is the Fourier transform of the pulse shaping filer sampled in $f=m\Delta f$.
The communication channel matrix can be expressed in the DD domain applying the inverse transformation in \eqref{eq:discreteISFFT}, obtaining \eqref{eq:comm_channel_DD}. 
Similarly, the sensing channel in \eqref{eq:channel_model_sensing} can be computed in the FT and DD domain, obtaining $\mathbf{H}_s^\mathrm{FT}$ and $\mathbf{H}_s^\mathrm{DD}$, by applying the same transformations.

While data symbols are estimated from \eqref{eq:TFreceivedsignal_sensing}, the sensing parameters (range and velocity of each UE) are retrieved in the DD domain as 
\begin{equation} \label{eq:DDreceivedsignal}
    \begin{split}
        \mathbf{Y}^{\mathrm{DD}}_s &= \mathbf{F}^{\mathrm{H}}_{M} \, \mathbf{Y}_s^{\mathrm{FT}} \, \mathbf{F}_{N} =\\
        & = \underbrace{\mathbf{F}^{\mathrm{H}}_{M}\widetilde{\mathbf{X}}^{\mathrm{FT}}_s\mathbf{F}_{N}}_{\text{Signal}} +
        \underbrace{\mathbf{F}^{\mathrm{H}}_{M}\widetilde{
\mathbf{X}}^{\mathrm{FT}}_c\mathbf{F}_{N} +\mathbf{N}_s^{\mathrm{DD}}}_{\text{Noise + Interference}},
    \end{split}
\end{equation}
where $\widetilde{\mathbf{X}}^{\mathrm{FT}}_s = \mathbf{H}^{\mathrm{FT}}_s\odot\mathbf{X}^{\mathrm{FT}}_s$ and $\widetilde{
\mathbf{X}}^{\mathrm{FT}}_c=\mathbf{H}^{\mathrm{FT}}_s\odot\mathbf{X}^{\mathrm{FT}}_c$. Due to the intrinsic property of the Fourier transform, the received signal $\mathbf{Y}_s^{\mathrm{DD}}$ is the convolution between the transmitted sensing signal $\mathbf{X}_s^{\mathrm{DD}}$ and the sensing channel $\mathbf{H}_s^\mathrm{DD}$. This means that the sensing impulse is shifted in the DD domain y a quantity equal to the delay and Doppler shift associated to UE's range and velocities, respectively.
Fig.~\ref{RX_DD} shows an example of received signal $\mathbf{Y}_s^{\mathrm{DD}}$ for $U=9$ UEs and random ranges and velocities.

\section{Numerical Results}\label{Simulations}
\begin{table}[t!]
    \centering
    \caption{Simulation Parameters}
    \begin{tabular}{l|c|c}
    \toprule
        \textbf{Parameter} &  \textbf{Symbol} & \textbf{Value(s)}\\
        \hline
        Carrier frequency & $f_c$ & $70$ GHz\\
        OFDM symbols (sensing)& $N$   & $128, 256, 512$\\
        OFDM subc. (sensing)& $M$       & $1024, 2049, 4096$\\
        OFDM symbols (comm.)& $N_{c,u}$   & $14$\\
        OFDM subc. (comm.)& $M_{c,u}$       & $240$\\
        Subc. spacing & $ \Delta f$       & $120$ kHz \\
        OFDM symbol time & $T$       & $8.9$ $\mu$s \\
        \bottomrule
    \end{tabular}
    \label{tab:SimParam}
\end{table}
To show the effectiveness of the proposed JC\&S waveform, we analyze the reciprocal impact of the communication on sensing and vice-versa.
In the considered scenario, the BS establishes a communication link with $U = 3$ UEs, located at range $R_1 = 15$ m, $R_2 = 25$ m and $R_3 = 35$ m, moving at radial velocities $V_1 = 14$ m/s, $V_2 = 25$ m/s and $V_3 = 30$ m/s, respectively. Only LoS paths are considered in both communication and sensing channels. Table \ref{tab:SimParam} reports the main simulation parameters.
\subsection{Communication Processing}
The metric used to evaluate the communication performance is the uncoded Bit Error Rate (BER) of a 16-QAM. We estimate the communication channel matrix $\mathbf{H}^{\mathrm{FT}}_{c,u}$ by inserting a pilot pattern following the NR 5G Demodulation Reference Signal (DMRS) specifications~\cite{ETSIphy}. After CP removal, the coefficients of $\mathbf{H}^{\mathrm{FT}}_{c,u}$ are first estimated over the pilot locations with a least squares approach, then DFT-interpolated along the columns and spline-interpolated along the rows~\cite{dong2007linear}, allowing the demodulation of the transmitted bits and the evaluation of the BER through Monte Carlo simulations.


The average downlink SNR over the FT resources, at the $u$-th UE, is
\begin{equation}\label{eq:SNR comm average}
        \gamma_{c,u}^{\mathrm{FT}}  = \frac{P_{c,u}^{\mathrm{FT}}\,G_{c,u}^{\mathrm{FT}}}{ P_s^{\mathrm{FT}}\,G_{c,u}^{\mathrm{FT}}+ P_{n,u}},
\end{equation}
where $G_{c,u}^{\mathrm{FT}}$ is the communication channel gain defined as
\begin{equation}\label{eq:average comm channel gain}
       G_{c,u}^{\mathrm{FT}} = \frac{1}{M_{c,u} N_{c,u}}\sum_{m \in \mathcal{M}_{c,u}}\sum_{n \in \mathcal{N}_{c,u}} \big\lvert H^{\mathrm{FT}}_{c,u}[m,n] \big\rvert^2.
\end{equation}
Note that the impact of the sensing signal on the communication performance depends on the sensing power in FT domain and not on the DD one.
\subsection{Sensing Processing}
The aim of sensing processing is to retrieve the physical parameters of the sensing channel $\mathbf{H}^{\mathrm{DD}}_s$ (delay and Doppler shift, i.e., range and velocity) from signal $\mathbf{Y}^{\mathrm{DD}}_s$. We estimate the position of the received peaks as 
\begin{align}
    \left[\widehat{\boldsymbol{\ell}},\widehat{\boldsymbol{k}}\right]= \underset{\boldsymbol{\ell},\boldsymbol{k}}{\mathrm{arg\, max}} \,\big\lvert Y^{\mathrm{DD}}_s[\ell,k]\big\rvert^2.
\end{align}
where we indicated with bold symbols $\boldsymbol{\ell},\boldsymbol{k}$ the selection of the first $U$ maxima of $Y^{\mathrm{DD}}_s[\ell,k]$ and we derive the vectors of estimated delay and Doppler shifts as
\begin{align}
    &\widehat{\boldsymbol{\tau}}=\Delta\tau\times[\widehat{\ell}_1-\ell_0,...,\widehat{\ell}_U-\ell_0]^{\mathrm{T}}\\
    &\widehat{\boldsymbol{\nu}}=\Delta\nu\times[\widehat{k}_1-k_0,...,\widehat{k}_U-k_0]^{\mathrm{T}}.
\end{align}
in which $[\ell_0,k_0]$ is the position of the Tx sensing pulse in DD domain. The resulting range and velocity estimates are then given by $\widehat{\boldsymbol{V}}= \widehat{\boldsymbol{\nu}} c/(2 f_c)$ $ \widehat{\boldsymbol{R}}=(c/2) \widehat{\boldsymbol{\tau}}$.

The sensing accuracy is expressed by computing the Root Mean Square Error (RMSE) on range and velocity:
\begin{align}\label{eq:RMSE}
    RMSE_{R} &= \frac{1}{U}\sqrt{\sum_{u=1}^{U}\mathbb{E}\left[\|\widehat{R}_u - R_u \|^2\right]},\\
    RMSE_{V} &= \frac{1}{U}\sqrt{\sum_{u=1}^{U}\mathbb{E}\left[\|\widehat{V}_u - V_u \|^2\right]}.
\end{align}
that are function of the average SNR over the DD lattice 
\begin{equation}\label{eq: av SNR sensing DD}
      \gamma_s^{\mathrm{DD}} = \frac{P_s^{\mathrm{DD}}G_s^{\mathrm{DD}}}{\sum_{u=1}^UP_{c,u}^{\mathrm{DD}}G_s^{\mathrm{DD}} + P_{n,s}},
\end{equation}
where $P_{n,s}$ is the additive noise power and $G_s^{\mathrm{DD}}$ is the sensing channel gain in DD domain
\begin{equation}
       G_{s}^{\mathrm{DD}} = \frac{1}{M N}\sum_{\ell = 0}^{M-1}\sum_{k = 0}^{N-1} \big\lvert H^{\mathrm{DD}}_{s}[\ell,k] \big\rvert^2.
\end{equation}
Similarly to \eqref{eq:SNR comm average}, the communication power associated to all the UEs can be seen as an additional noise in DD domain.
\subsection{Results}
%

\begin{figure}[!t] 
    \centering
    \vspace{0.45 cm}\includegraphics[width=0.66\columnwidth]{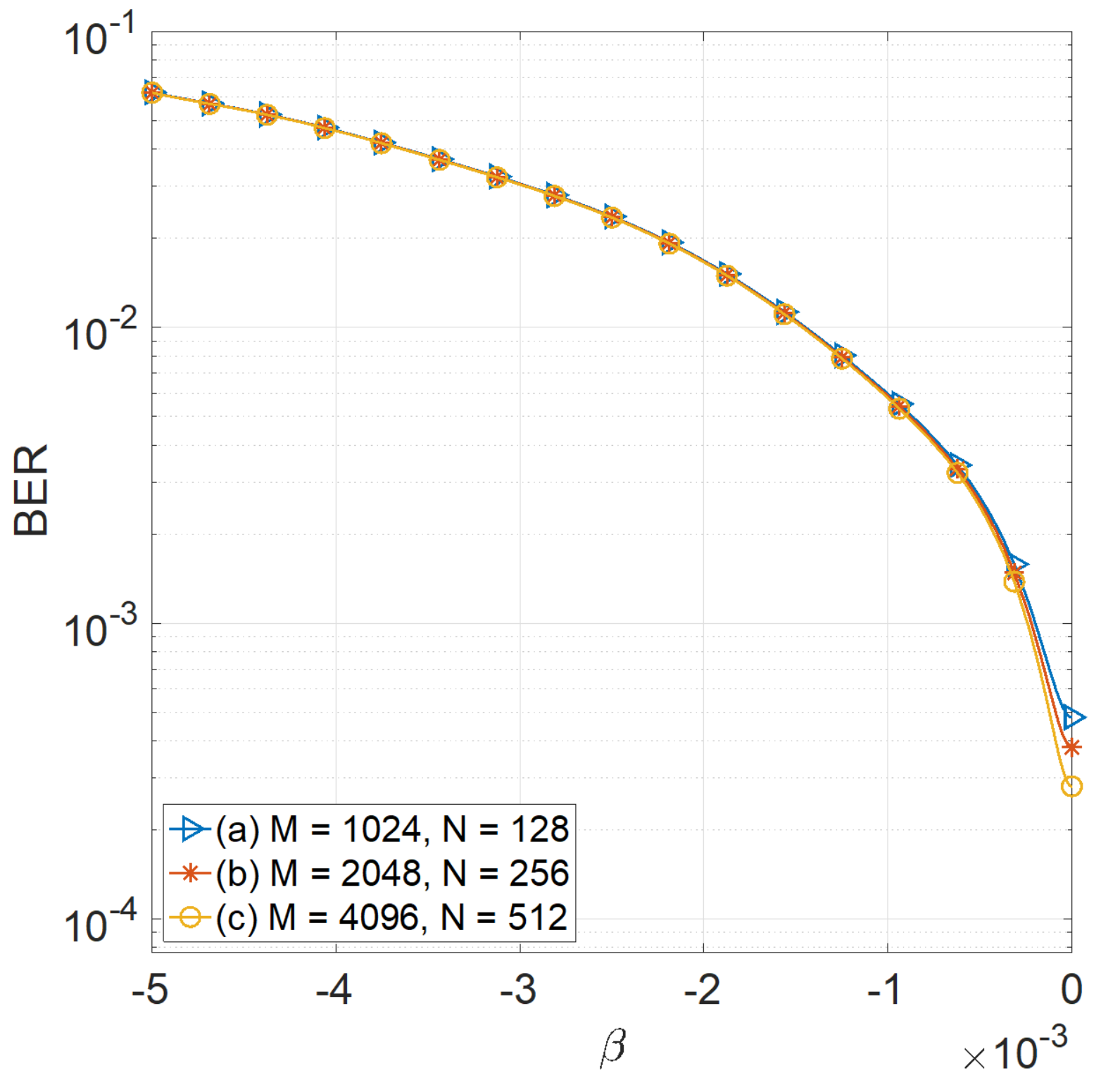}
    \caption{Average BER vs $\beta$ for different FT resources, $M$ = $1024$ and $N$ = $128$ (blue), $M$ = $2048$ and $N$ = $2565$ (red), and $M$ = $4096$ and $N$ = $512$ (yellow).}
    \label{fig:BER}
\end{figure}

\begin{figure}[!t] 
    \centering
    \vspace{0.45 cm}
    \subfloat[][]{\includegraphics[width=0.66\columnwidth]{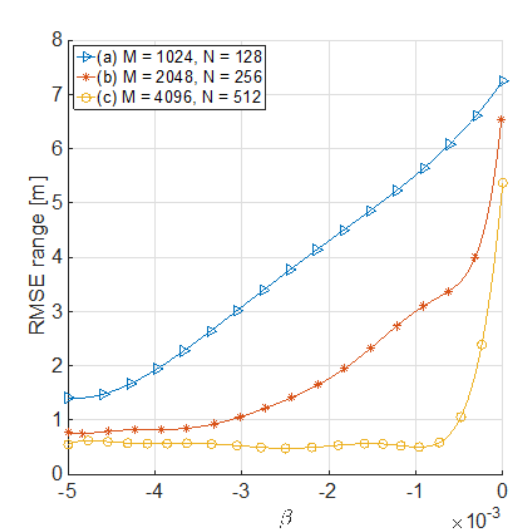}\label{fig:rangeRMSE}}\\
    \subfloat[][]{\includegraphics[width=0.66\columnwidth]{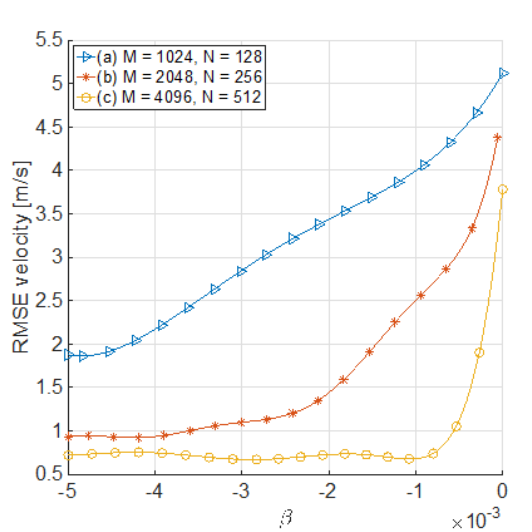} \label{fig:velRMSE}}
\caption{Average RMSE vs $\beta$ for different FT resources, $M$ = $1024$ and $N$ = $128$ (blue), $M$ = $2048$ and $N$ = $2565$ (red), and $M$ = $4096$ and $N$ = $512$ (yellow). Range (top) and velocity (bottom).}
\label{fig:RMSE}
\end{figure}

This section highlights the simulation results by considering 5G NR numerology $\mu = 3$~\cite{3gpp.38.211}, therefore a subcarrier spacing of $\Delta f=120$ kHz and an OFDM symbol duration of $T=8.9$ $\mu$s. We aim at showing that the proposed waveform allows to properly perform both communication and sensing functionalities.
In order to show the reciprocal impact of communication and sensing signal, we show the BER and RMSE performance, varying the power allocated to each waveform ($P_{c}^{\mathrm{FT}}$ and $P_{s}^{\mathrm{FT}}$). Specifically, we fix the total power budget to $P_{tot}^{\mathrm{FT}} = 20$ mW ($13$ dBm), and determine the communication and sensing Tx powers as
\begin{equation}
    P_{c}^{\mathrm{FT}} = \rho P_{tot}^{\mathrm{FT}},\,\,\,\,P_{s}^{\mathrm{FT}} = (1-\rho)P_{tot}^{\mathrm{FT}},
\end{equation}
where $0 < \rho < 1$. In practice, the working condition of the proposed JC\&S system is for values of $\rho$ very close to 1, i.e., $\rho = 10^\beta$, with $\beta \in [-5\cdot10^{-3},\dots,0)$, as shown in Figs.~\ref{fig:BER} and~\ref{fig:RMSE}.  

%
Figure~\ref{fig:BER} reports the impact of the sensing waveform in \eqref{eq:discreteISFFT} on the received OFDM symbols with communication Tx power $P_{c,u}^{\mathrm{FT}}=10^{\beta}P_{tot}^{\mathrm{FT}}$, $u$\,=\,$1,\dots,U$ varying $\beta$. We take into account three different FT sensing allocated resources, namely (a) $M$ = $1024$ and $N$ = $128$ (blue), (b) $M$ = $2048$ and $N$ = $2565$ (red), and (c) $M$ = $4096$ and $N$ = $512$ (yellow). By increasing $\beta$, the BER decreases without being affected by $M$ and $N$, since the communication performance in \eqref{eq:SNR comm average} is dominated by the communication channel gain $G_{c,u}^{\mathrm{FT}}$ in \eqref{eq:average comm channel gain}.  With a proper power allocation, therefore, the sensing power does not affect communication.

The range and velocity RMSEs are shown in Figs.~\ref{fig:rangeRMSE} and~\ref{fig:velRMSE}, respectively, varying $\beta$ in the same range of Fig.~\ref{fig:BER}, for the same values of $M$ and $N$. We observe that for comparatively small $\beta$ values (i.e., a higher Tx sensing power) the system correctly reaches the range and velocity resolution bounds, corresponding to (a) $\Delta R \approx 1.2$ m, $\Delta V \approx 1.9$ m/s (b) $\Delta R \approx 0.6$ m, $\Delta V \approx 0.9$ m/s (c) $\Delta R \approx 0.3$ m, $\Delta V \approx 0.46$ m/s. An increase in $M$ and $N$ means an increase of $P_s^{\mathrm{DD}}$ given a fixed $P_s^{\mathrm{FT}} = (1-10^\beta)P_{tot}^{\mathrm{FT}}$, as from \eqref{eq:power_sensing_FT}, leading to enhanced sensing performance for a fixed BER value. In these settings, by selecting $\beta\in[-1.5,-0.5]\cdot10^{-3}$, we can achieve a BER $\leq10^{-2}$, ensuring a sensing performance approaching the resolution bound for $M=4096$, $N=512$ (case (c)). This results suggests that, for a proper power budget (mostly allocated to communication), it is possible to find a working condition for both functionalities with a full FT resource sharing between the two. For a fixed target BER, the quality of the sensing information, in terms of both resolution and RMSE on range and velocity, is set by the choice of $M$ and $N$, provided that the values are within practical limits, i.e., maximum available bandwidth or length of the downlink burst. It is worth noticing that, provided that the necessary power budget $P^{\mathrm{FT}}_{tot}$ is within the typical capabilities of the JC\&S transmitter (e.g., $43$ dBm for the BS), the proposed system does not set an explicit trade-off between communication and sensing parameters, making it a low-complexity solution for future 6G systems.



%
\section{Conclusion}\label{Conclusion}
In this paper, we propose a novel JC\&S waveform to enable sensing capabilities on top of a communication system, based on the superposition of a properly designed signal in the DD domain and an OFDM communication signal in the FT domain. The dual domain JC\&S waveform is first analytically described, pointing out the relations of the signals in different domains and their mutual interference. The proposed waveform is able to carry information to multiple UEs while sensing the environment, estimating range and velocity of the UEs.
Numerical results show that, with a fixed power budget and a proper power allocation to communication and sensing, the capabilities of the latter (both resolution) can be tuned by the choice of the available bandwidth and length of the downlink burst, while leaving the BER substantially unaffected.

%

\section*{Acknowledgment}
The work has been partially framed within the Huawei-
Politecnico di Milano Joint Research Lab. 

\bibliographystyle{IEEEtran}
\bibliography{Bibliography}

\end{document}